\documentclass[12pt]{article}		
\usepackage[english]{babel} 		
\usepackage{fullpage}				
\usepackage[utf8]{inputenc}			
\usepackage{amsmath, amssymb,amsthm}
\usepackage[final]{graphicx}				
\usepackage[colorinlistoftodos]{todonotes}	
\usepackage{tikz}
\usepackage{braket}
\usepackage{physics}
\usepackage{bm}
\usepackage{float}
\usepackage{slashed}
\usepackage{abstract}
\usepackage{authblk}
\usepackage{hyperref}
\usepackage{cite}
\usepackage{indentfirst}

\usepackage{mathdots}
\usepackage{yhmath}
\usepackage{cancel}
\usepackage{color}
\usepackage{siunitx}
\usepackage{array}
\usepackage{multirow}
\usepackage{amssymb}
\usepackage{gensymb}
\usepackage{tabularx}
\usepackage{booktabs}
\usetikzlibrary{fadings}
\usetikzlibrary{patterns}
\usetikzlibrary{shadows.blur}
\usetikzlibrary{shapes}

\newcommand{\half}{\frac{1}{2}}

\title{{\bf Reconstructing the Bulk Dual of ABJM from Holographic Entanglement Entropy}}

\author{Ashton Lowenstein\thanks{alowenst@usc.edu} }
\author{Avik Chakraborty\thanks{avikchak@usc.edu}}

\affil{{\it Department of Physics and Astronomy} \\ {\it University of Southern California} \\ {\it Los Angeles, CA 90089, USA}}

\date{\vspace{-5ex}}

\begin{document}
\maketitle
\pagenumbering{gobble}

\thispagestyle{empty}

\begin{abstract}

Recent work has shown that entanglement and the structure of spacetime are intimately related. One way to investigate this is to begin with an entanglement entropy in a conformal field theory (CFT) and use the AdS/CFT correspondence to calculate the bulk metric. We perform this calculation for ABJM, a particular 3-dimensional supersymmetric CFT (SCFT), in its ground state. In particular we are able to reconstruct the pure $AdS_4$ metric from the holographic entanglement entropy of the boundary ABJM theory in its ground state. Moreover, we are able to predict the correct AdS radius purely from entanglement. We also address the general philosophy of relating entanglement and spacetime through the Holographic Principle, as well as some of the philosophy behind our calculations.

\end{abstract}

\clearpage
\pagenumbering{arabic}
\tableofcontents

\section{Introduction}

\indent The discovery of the AdS/CFT correspondence \cite{Maldacena:1997re, Aharony:1999ti, Witten:1998qj, Horowitz:2006ct} has proven to be one of the most fruitful and interesting events in modern theoretical physics. The correspondence provides a realization of the Holographic Principle \cite{Susskind:1994vu, tHooft:1993dmi}, which states that the degrees of freedom of a `bulk' theory can be described by degrees of freedom on its boundary. Another attractive feature of the correspondence is its connection between strongly coupled field theory and weakly coupled gravity.

In its most basic form, the correspondence relates quantities computable in a theory of gravity, specifically a spacetime which is asymptotically AdS, to ones calculated in a `dual' quantum field theory through what's called a `holographic dictionary.' For example, the behavior of scalar fields in the bulk gravity theory can be determined by studying correlation functions of operators in the boundary field theory \cite{Witten:1998qj}. The metric of the bulk gravity theory is also determined in part by the stress tensor of the dual conformal field theory (CFT) \cite{deHaro:2000vlm}. In \cite{Ryu:2006bv, Ryu:2006ef} Ryu and Takayanagi proposed a relationship between entanglement entropy in the dual field theory and the area of an extremal surface in the bulk gravity theory. 

In particular, suppose the field theory is defined on a manifold $\mathcal{M}$ which has the topology $\mathbb{R} \times \Sigma$, where $\Sigma$ is a spacelike hypersurface. Let $\rho$ be a global pure state in the field theory and $A \subset \Sigma$ a constant-time subregion of $\mathcal{M}$. The boundary $\partial A$ is called the entangling surface. The reduced density matrix $\rho_A$ is defined as the partial trace of $\rho$ over degrees of freedom in the complement $A^C$. The entanglement entropy of the state $\rho_A$ is then defined as
\begin{equation}
S_A \equiv S[\rho_A] = -\tr_A(\rho_A\log \rho_A).
\end{equation}

The dual spacetime geometry will be an asymptotically AdS manifold whose conformal boundary is $\mathcal{M}$. Let $\Gamma$ be the constant-time surface which is anchored along $\partial A$ on the boundary with the extremal volume $\mathcal{A}[\Gamma]$ (calculated with the bulk metric). Then the Ryu-Takayanagi (RT) formula states that
\begin{equation}
S_A = \frac{\mathcal{A}[\Gamma]}{4G_N}. \label{eqn:RT}
\end{equation}

In practice it is easiest to use this formula to compute the entanglement entropy for a CFT in some global state (usually the vacuum or a thermal state) reduced to a region $A$. That is, one gains information about the boundary field theory by performing a calculation using geometric information in the bulk. Recent studies of entanglement entropy in the context of AdS/CFT indicate that entanglement is intimately related to the structure of spacetime and gravity theories \cite{Maldacena:2013xja, VanRaamsdonk:2010pw, Dong:2016eik, Cotler:2017erl, Penington:2019npb, Chen:2019gbt, Bilson:2008ab, Bilson:2010ff, Faulkner:2014jva}. From this perspective, one might ask how much can be learned about the bulk geometry entanglement in the CFT. This question has been explored in several ways, including the ER = EPR proposal, entanglement wedge reconstruction with and without using the Petz map, reconstructing bulk geodesics, and reconstructing the bulk metric.

To philosophically motivate the calculations later in the paper, we can begin by summarizing an argument in \cite{VanRaamsdonk:2010pw}. As a thought experiment, picture a field theory defined on a sphere $S^2$, and split this sphere into two regions $A,B$. The extremal surface anchored along the boundary of $A$ is a sheet that splits the bulk ball into two distinct regions $\tilde{A},\tilde{B}$. Now suppose we can calculate the entanglement entropy of some state $\rho$ reduced to $A$, and suppose that the entanglement entropy is tuneable with some parameter. That is, if we increase the parameter the entanglement goes up, and vice versa. Imagine we tune the entanglement entropy down to 0. By the RT formula, the area of of the bulk extremal surface will also shrink to 0, meaning the regions $\tilde{A},\tilde{B}$ become totally disjoint. The effect of having no entanglement entropy between the two regions in the CFT is to have two separate spacetimes. In this sense, the bulk spacetime is held together by entanglement entropy. A cartoon of this process in one less dimension is presented below in fig. (\ref{fig:mvr}).

\begin{figure}[!ht]
\center

\tikzset{every picture/.style={line width=0.75pt}} 

\begin{tikzpicture}[x=0.75pt,y=0.75pt,yscale=-1,xscale=1]

\draw   (2.22,83.03) .. controls (2.22,49.32) and (29.54,22) .. (63.25,22) .. controls (96.96,22) and (124.28,49.32) .. (124.28,83.03) .. controls (124.28,116.74) and (96.96,144.06) .. (63.25,144.06) .. controls (29.54,144.06) and (2.22,116.74) .. (2.22,83.03) -- cycle ;
\draw [color={rgb, 255:red, 208; green, 2; blue, 27 }  ,draw opacity=1 ]   (63.25,144.06) .. controls (35.88,110.77) and (42.91,44.19) .. (63.25,22) ;
\draw   (126.68,79.7) -- (151.31,79.7) -- (151.31,72.3) -- (167.73,87.1) -- (151.31,101.89) -- (151.31,94.5) -- (126.68,94.5) -- cycle ;
\draw  [draw opacity=0] (252.18,113.43) .. controls (244.05,126.58) and (230.6,135.18) .. (215.37,135.18) .. controls (190.55,135.18) and (170.43,112.33) .. (170.43,84.14) .. controls (170.43,55.95) and (190.55,33.1) .. (215.37,33.1) .. controls (230.04,33.1) and (243.07,41.08) .. (251.27,53.43) -- (215.37,84.14) -- cycle ; \draw   (252.18,113.43) .. controls (244.05,126.58) and (230.6,135.18) .. (215.37,135.18) .. controls (190.55,135.18) and (170.43,112.33) .. (170.43,84.14) .. controls (170.43,55.95) and (190.55,33.1) .. (215.37,33.1) .. controls (230.04,33.1) and (243.07,41.08) .. (251.27,53.43) ;
\draw [color={rgb, 255:red, 208; green, 2; blue, 27 }  ,draw opacity=1 ]   (251.27,53.43) .. controls (236.5,61) and (231.37,108.02) .. (250.85,113.88) ;
\draw   (335.2,78.7) -- (359.84,78.7) -- (359.84,71.3) -- (376.26,86.1) -- (359.84,100.89) -- (359.84,93.5) -- (335.2,93.5) -- cycle ;
\draw   (431.43,103.84) .. controls (419.46,115.42) and (400.38,115.1) .. (388.8,103.13) .. controls (377.23,91.16) and (377.55,72.08) .. (389.52,60.5) .. controls (401.49,48.93) and (420.57,49.25) .. (432.14,61.22) .. controls (446.11,75.66) and (453.1,82.89) .. (453.1,82.89) .. controls (453.1,82.89) and (445.88,89.88) .. (431.43,103.84) -- cycle ;
\draw   (473.93,61.1) .. controls (473.93,61.1) and (473.93,61.1) .. (473.93,61.1) .. controls (485.43,49.06) and (504.52,48.63) .. (516.55,60.14) .. controls (528.59,71.64) and (529.02,90.72) .. (517.51,102.76) .. controls (506.01,114.79) and (486.92,115.22) .. (474.89,103.72) .. controls (460.36,89.83) and (453.09,82.89) .. (453.1,82.89) .. controls (453.09,82.89) and (460.04,75.62) .. (473.93,61.1) -- cycle ;
\draw  [color={rgb, 255:red, 208; green, 2; blue, 27 }  ,draw opacity=1 ][line width=3] [line join = round][line cap = round] (453.94,82.36) .. controls (453.94,76.28) and (448.12,88.17) .. (453.94,82.36) ;
\draw  [color={rgb, 255:red, 208; green, 2; blue, 27 }  ,draw opacity=1 ][line width=3] [line join = round][line cap = round] (452.46,83.1) .. controls (452.46,85.07) and (454.67,83.19) .. (454.67,82.36) ;
\draw   (459.74,109.07) -- (460.1,133.7) -- (467.49,133.59) -- (452.94,150.23) -- (437.91,134.03) -- (445.3,133.92) -- (444.94,109.29) -- cycle ;
\draw   (375.15,180.01) .. controls (375.15,159.99) and (391.38,143.76) .. (411.4,143.76) .. controls (431.42,143.76) and (447.65,159.99) .. (447.65,180.01) .. controls (447.65,200.03) and (431.42,216.25) .. (411.4,216.25) .. controls (391.38,216.25) and (375.15,200.03) .. (375.15,180.01) -- cycle ;
\draw   (458,179.27) .. controls (458,159.25) and (474.23,143.02) .. (494.25,143.02) .. controls (514.27,143.02) and (530.5,159.25) .. (530.5,179.27) .. controls (530.5,199.29) and (514.27,215.52) .. (494.25,215.52) .. controls (474.23,215.52) and (458,199.29) .. (458,179.27) -- cycle ;
\draw  [draw opacity=0] (250.99,53.87) .. controls (259.35,40.86) and (272.96,32.5) .. (288.18,32.77) .. controls (313,33.21) and (332.71,56.41) .. (332.21,84.6) .. controls (331.72,112.78) and (311.2,135.28) .. (286.38,134.84) .. controls (271.71,134.58) and (258.83,126.37) .. (250.85,113.88) -- (287.28,83.81) -- cycle ; \draw   (250.99,53.87) .. controls (259.35,40.86) and (272.96,32.5) .. (288.18,32.77) .. controls (313,33.21) and (332.71,56.41) .. (332.21,84.6) .. controls (331.72,112.78) and (311.2,135.28) .. (286.38,134.84) .. controls (271.71,134.58) and (258.83,126.37) .. (250.85,113.88) ;

\draw (109.76,21.54) node [anchor=north west][inner sep=0.75pt]   [align=left] {$\displaystyle A$};
\draw (4.18,21.8) node [anchor=north west][inner sep=0.75pt]   [align=left] {$\displaystyle B$};
\draw (77.25,64.53) node [anchor=north west][inner sep=0.75pt]   [align=left] {$\displaystyle \tilde{A}$};
\draw (309.64,22.2) node [anchor=north west][inner sep=0.75pt]   [align=left] {$\displaystyle A$};
\draw (174.84,22.06) node [anchor=north west][inner sep=0.75pt]   [align=left] {$\displaystyle B$};
\draw (53.62,36.02) node [anchor=north west][inner sep=0.75pt]   [align=left] {$\displaystyle \textcolor[rgb]{0.82,0.01,0.11}{RT}$};
\draw (238.38,68.86) node [anchor=north west][inner sep=0.75pt]   [align=left] {$\displaystyle \textcolor[rgb]{0.82,0.01,0.11}{RT}$};
\draw (485.51,29.8) node [anchor=north west][inner sep=0.75pt]   [align=left] {$\displaystyle A$};
\draw (401.18,29.8) node [anchor=north west][inner sep=0.75pt]   [align=left] {$\displaystyle B$};
\draw (441.31,53.91) node [anchor=north west][inner sep=0.75pt]   [align=left] {$\displaystyle \textcolor[rgb]{0.82,0.01,0.11}{RT}$};
\draw (400.92,122.27) node [anchor=north west][inner sep=0.75pt]   [align=left] {$\displaystyle B$};
\draw (485.99,120.79) node [anchor=north west][inner sep=0.75pt]   [align=left] {$\displaystyle A$};
\draw (18,65) node [anchor=north west][inner sep=0.75pt]   [align=left] {$\displaystyle \tilde{B}$};
\draw (198,65) node [anchor=north west][inner sep=0.75pt]   [align=left] {$\displaystyle \tilde{B}$};
\draw (404,65) node [anchor=north west][inner sep=0.75pt]   [align=left] {$\displaystyle \tilde{B}$};
\draw (404,159) node [anchor=north west][inner sep=0.75pt]   [align=left] {$\displaystyle \tilde{B}$};
\draw (281.25,64.53) node [anchor=north west][inner sep=0.75pt]   [align=left] {$\displaystyle \tilde{A}$};
\draw (485.25,64.53) node [anchor=north west][inner sep=0.75pt]   [align=left] {$\displaystyle \tilde{A}$};
\draw (485.25,158.53) node [anchor=north west][inner sep=0.75pt]   [align=left] {$\displaystyle \tilde{A}$};

\end{tikzpicture}
\caption{A cartoon demonstrating how tuning the entanglement entropy to zero produces two disjoint spacetimes. The RT surface is shown in red.}
\label{fig:mvr}
\end{figure}
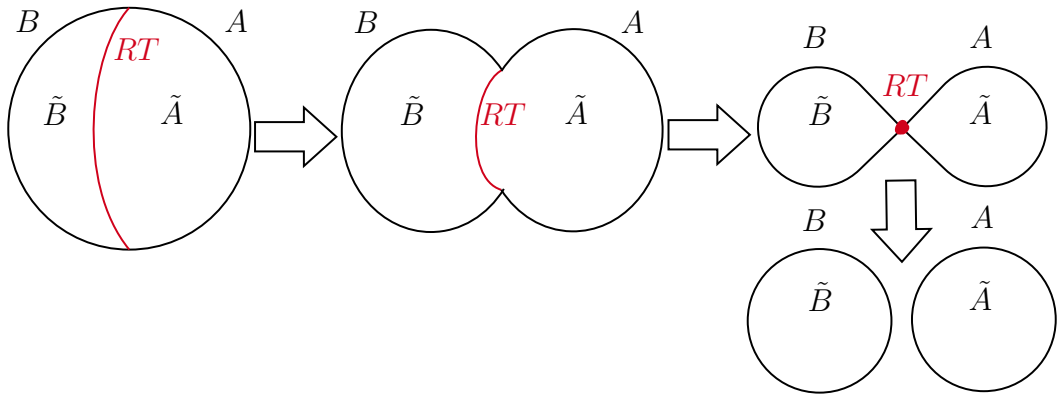

The rest of the paper is organized as follows. In section 2 we present a qualitative argument demonstrating the relationship between entanglement and spacetime. In section 3 we review the basics of ABJM \cite{Aharony:2008ug} and present results for the entanglement entropy of several entangling surfaces. We argue that when the entangling surface is a finite length rectangular strip the entanglement entropy should match the RT result. In section 4 we introduce the framework for computing the bulk metric from entanglement entropy and perform the calculation for ABJM in Minkowski space. We successfully reconstruct the pure $AdS_4$ metric with the correct AdS radius from the entanglement entropy of the ABJM vacuum reduced to a strip in Minkowski space. Examples of this methodology for 2-dimensional CFTs in various states and $\mathcal{N} = 4$ SYM in the vacuum were explored in \cite{Saha:2018jjb}.

\section{ABJM and AdS/CFT}

\indent The ABJM theory \cite{Aharony:2008ug} is a three dimensional supersymmetric Chern-Simons-matter theory with gauge group $U(N)_k \times U(N)_{-k}$ where $(k,-k)$ are the Chern-Simons levels of the action. The theory is superconformal, generally with $\mathcal{N} = 6$ supersymmetry, and is weakly coupled in the limit $k \gg N$. The supersymmetry is enhanced to $\mathcal{N} = 8$ when $N = 2$, as well as when $k = 1,2$.

The theory can be constructed by considering a coincident stack of M2-branes in 11-dimensional supergravity. When the stack of $N$ branes is probing a $\mathbb{C}^4/\mathbb{Z}_k$ singularity, the field theory description of the branes is an $\mathcal{N} = 3$ predecessor of ABJM, namely a supersymmetric Chern-Simons-matter theory with massive dynamical gauge fields. One can integrate out the gauge fields in the low energy limit to recover the $\mathcal{N} = 6$ ABJM theory. In the limit of large $N$ this brane construction can be thought of M-theory on $AdS_4/\mathbb{Z}_k$. From this perspective, the Chern-Simons level $k$ is related to the circle one would dimensionally reduce M-theory on to get type IIA string theory. Thus this pairing of $AdS_4$ and ABJM gives a realization of the AdS/CFT correspondence.

There has been a great deal of success calculating the partition function for ABJM on (squashed and branched) 3-spheres using localization techniques (for a review, see \cite{Marino:2011nm}). The partition function of ABJM on $S^3$ can then be used to calculate entanglement entropy for ABJM in a different context. Using conformal transformations, it can be shown that \cite{Casini:2011kv} the entanglement entropy of the vacuum reduced to a disk is given by
\begin{equation}
S_{\text{disk}} = \log Z_{S^3}, \label{eqn:EEpartitionfunction}
\end{equation}
where $Z_{S^3}$ is the $S^3$ partition function. To leading order in $N$ the universal part of the result is \cite{Drukker:2010nc}
\begin{equation}
\log Z_{S^3} = \frac{\sqrt{2}\pi}{3} k^{1/2}N^{3/2}.
\end{equation}
Including the divergent area term and choosing the UV regulator appropriately, the entanglement entropy becomes \cite{Ryu:2006bv}
\begin{equation}
S_{\text{disk}} = \frac{\sqrt{2}\pi}{3} k^{1/2}N^{3/2}\left( \frac{\ell}{a} - 1\right),
\end{equation}
where $\ell$ is the radius of the disk and $a$ is the UV regulator. This is expected for a 3-dimensional CFT (see {\it e.g.} \cite{Klebanov:2011td} and references therein).

The construction of Casini {\it et al.} \cite{Casini:2011kv} which make the entanglement entropy calculation tractable on the field theory side rely heavily on the spherical symmetry of the entangling surface. However, using a spherical entangling surface renders the bulk metric reconstruction calculation very difficult. It will be much easier to use a rectangular strip as our entangling surface. In fact, it is possible to calculate the entanglement entropy of the vacuum reduced to a rectangular strip using holography \cite{Ryu:2006bv}. If the strip has width $\ell$ and length $L$ (understood to be very large), the result is
\begin{equation}
S_{\text{strip}} = \frac{\sqrt{2}}{3} k^{1/2}N^{3/2} \left(\frac{L}{a} - \frac{ 4\pi^3L }{ \Gamma(1/4)^4\ell } \right). \label{eqn:Sstrip}
\end{equation}

\section{Bulk metric reconstruction}

\indent The general philosophy for using entanglement entropy to calculate the metric of the bulk spacetime is as follows. Suppose we start with just the idea of a holographic dictionary. In particular, putting the boundary CFT in its vacuum state should be dual to a relatively `calm' bulk geometry. This is exemplified by Fefferman-Graham coordinates \cite{AST_1985__S131__95_0, Fefferman2007TheAM}, in the sense that the geometry is affected by the expectation value of the stress tensor of the field theory. Further, we want the conformal boundary of our bulk spacetime to match the spacetime the CFT is defined on. Along with this, we are free to use diffeomorphism invariance of the bulk gravity theory to choose coordinates which are convenient. The final part of the holographic dictionary that we need is the RT formula (\ref{eqn:RT}). Using these principles we write down an ansatz for the bulk metric.

\begin{figure}[!ht]
\center
	\includegraphics[scale=1]{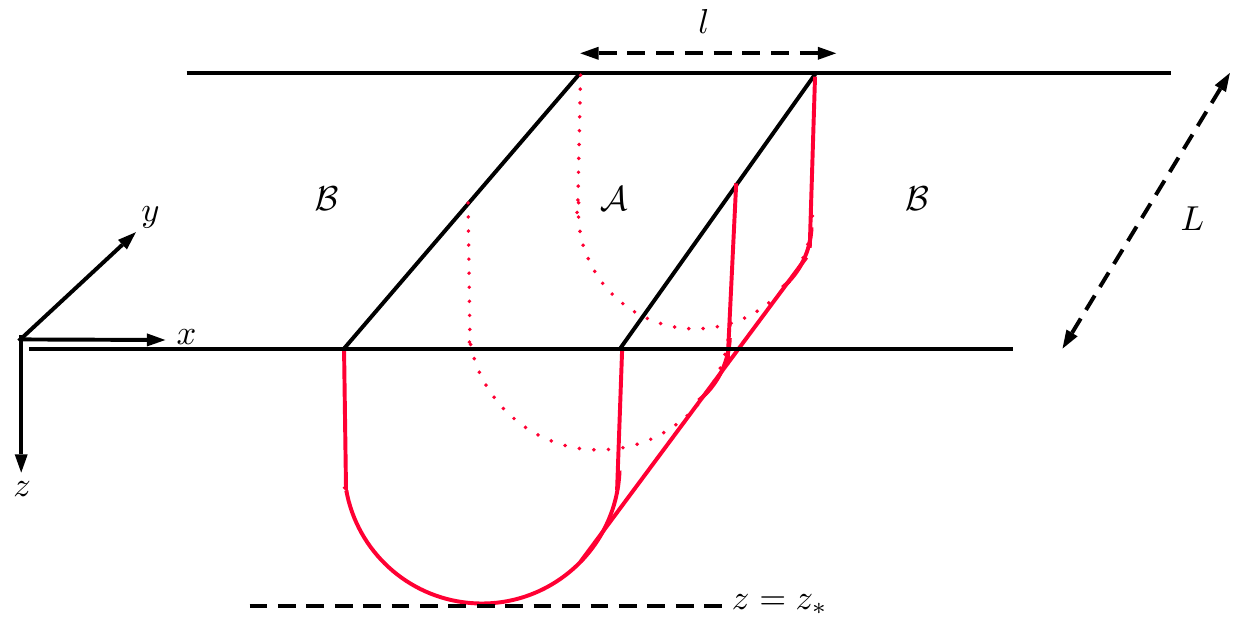}
	\caption{A sketch of a the entangling region in a 3d CFT and its corresponding RT surface in the bulk.}
	\label{fig:stripgeom}
\end{figure}

For the methodology of the calculation we follow \cite{Bilson:2008ab, Bilson:2010ff}. Consider a CFT defined on $d$-dimensional Minkowski space. Suppose we know the entanglement entropy of this theory's vacuum reduced to a rectangular strip. Schematically, we expect that since the bulk geometry is not too perverse ({\it i.e.} calm), the RT surface should share some of the symmetry of the entangling surface. As such, we choose the metric ansatz
\begin{equation}
ds^2 = \frac{R^2}{z^2}\left(-h(z) dt^2 + f(z)^2dz^2 + \sum_{i = 1}^{d-2} dx_i^2 \right).
\end{equation}
Arrange the entangling region to be
\begin{equation}
A:\quad \Big\{x_i\,|\, -\ell/2 < x_1 < \ell/2, \quad 0< x_2,x_3,\dots,x_{d-2} < L \Big\}.
\end{equation}
An example is shown in fig. \ref{fig:stripgeom}.

The symmetry between $x_2,\dots,x_{d-2}$ informs us to assume the RT surface can be parametrized in the bulk by $z = z(x_1)$. The induced metric on the RT surface is
\begin{equation}
ds^2 = \frac{R^2}{z^2} \left( \Big(z'^2f^2 + 1\Big)dx_1^2 + \sum_{i = 2}^{d-2}dx_i^2 \right),
\end{equation}
where $z' \equiv dz/dx$. Call the RT surface $\gamma$. The volume of $\gamma$ is
\begin{equation}
\mathcal{A}[\gamma] = R^{d-1}L^{d-2} \int_{\ell/2}^{\ell/2} dx\,\frac{\sqrt{1 + z'(x)^2f(z(x))^2}}{z(x)^{d-1}}, \label{eqn:genarea}
\end{equation}
where we have integrated out the extra coordinates. This volume is a functional of the embedding $z(x)$ and the metric function $f(z)$. 

According to the RT prescription, we must extremize (\ref{eqn:genarea}) with respect to $f,z$. We can use the calculus of variations to solve the problem. Call the integrand the Lagrangian
\begin{equation}
\mathcal{L}(z,z';x) = \frac{\sqrt{1 + (z'f(z))^2}}{z^{d-1}},
\end{equation}
where $z,z'$ are the generalized coordinates and $x$ is the `time'. Since $\mathcal{L}$ does not depend explicitly on $x$, its Legendre transform (the Hamiltonian) is conserved
\begin{equation}
\mathcal{H} = z' \frac{\partial \mathcal{L}}{\partial z'} - \mathcal{L},\quad\quad \frac{d\mathcal{H}}{dx} = 0.
\end{equation}
The Hamiltonian is given by
\begin{equation}
\mathcal{H} = -\frac{1}{z^{d-1}\sqrt{1 + (z'f)^2}}.
\end{equation}
Because the Hamiltonian is conserved, we can set it equal to something that does not depend on $x$. In particular, we choose to relate it to the turning point of the RT surface, called $z_\ast$:
\begin{equation}
\mathcal{H} = -\frac{1}{z_\ast^{d-1}}.
\end{equation}

The next steps, which will be displayed specifically for the case of interest below, are to rewrite the volume functional using $z_\ast$, relate the field theory entanglement entropy to the area functional using the RT formula, and then solve the resulting integral equation for the metric function $f(z)$. Note that this procedure does not give us the time component of the metric $h(z)$.

\subsection{The bulk dual of ABJM in its ground state}

\indent Write the metric ansatz
\begin{equation}
ds^2 = \frac{R^2}{z^2}\Big(-h(z)dt^2 + f(z)^2dz^2 + dx^2 + dy^2\Big).
\end{equation}
Assume the RT surface is given by $z = z(x)$ with $x \in (-\ell/2,\ell/2)$ and $y \in (0,L)$. Since the RT surface $\gamma$ is 2-dimensional, we will start referring to $\mathcal{A}$ as the area. The area functional is
\begin{equation}
\mathcal{A} 
= 2LR^2 \int_0^{\ell/2}dx\,\frac{\sqrt{1 + (z'f(z))^2}}{z^2}.
\end{equation}
In the calculus of variations problem the Lagrangian is
\begin{equation}
\mathcal{L}(z,z';x) = \frac{\sqrt{1 + (z'f(z))^2}}{z^2},
\end{equation}
and the conserved Hamiltonian is
\begin{equation}
H 
= -\frac{1}{z^2\sqrt{1 + (z'f)^2}}.
\end{equation}
Let $z_\ast$ be the turning point of the surface, so that $\frac{dz}{dx}\Big|_{z = z_\ast} = 0$. Then
\begin{equation}
z' = \frac{\sqrt{z_\ast^4 - z^4}}{z^2f(z)},
\end{equation}
which implies
\begin{equation}
\ell = 2 \int_a^{z_\ast} dz\,\frac{z^2f(z)}{\sqrt{z_\ast^4 - z^4}},
\end{equation}
where $a$ is a UV cutoff. In terms of $z_\ast$ the Lagrangian is $\mathcal{L} = z_\ast^2/z^4$,
and our area functional becomes
\begin{equation}
\mathcal{A} = 2LR^2 \int_a^{z_\ast} dz\,\frac{z_\ast^2 f(z)}{z^2 \sqrt{z_\ast^4 - z^4}}, \label{eqn:areafunctional}
\end{equation}

Next, we view the entanglement entropy in eqn. (\ref{eqn:Sstrip}) and the area functional in eqn. (\ref{eqn:areafunctional}) as functions of the parameter $\ell$. Their $\ell$ derivatives are 
\begin{align}
\frac{dS_{\text{strip}}}{d\ell} &=  \frac{\sqrt{2}}{3}k^{1/2}N^{3/2}\frac{4 \pi^3L}{\Gamma(1/4)^4 \ell^2}, \\
\frac{d\mathcal{A}}{d\ell} &= \frac{dA}{dz_\ast}\frac{dz_\ast}{d\ell}.
\end{align}
Because $\mathcal{A}$ does not explicitly depend on $\ell$ we must use the chain rule. Write
\begin{equation}
\mathcal{A} = 2LR^2 \int_a^{z_\ast} F(z_\ast,z)dz \quad \& \quad \ell = 2 \int_a^{z_\ast} \frac{z^4}{z_\ast^2}F(z_\ast,z)dz,
\end{equation}
where
\begin{equation}
F(z_\ast,z) = \frac{z_\ast^2 f(z)}{z^2 \sqrt{z_\ast^4 - z^4}}.
\end{equation}
By the Leibniz rule for differentiating an integral
\begin{align*}
\frac{d\mathcal{A}}{dz_\ast} 
&= 2LR^2 \left[ \lim_{z \to z_\ast} F(z_\ast,z) -2 \int_a^{z_\ast} \frac{z_\ast z^2 f(z)}{\left(z_\ast^4 - z^4\right)^{3/2}}dz \right] \\
\frac{d\ell}{dz_\ast} 
&= 2z_\ast^2  \left[ \lim_{z \to z_\ast} F(z_\ast,z) -2 \int_a^{z_\ast} \frac{z_\ast z^2 f(z)}{\left(z_\ast^4 - z^4\right)^{3/2}}dz \right].
\end{align*}
Hence
\begin{equation}
\frac{d\mathcal{A}}{d\ell} = \frac{LR^2}{z_\ast^2}.
\end{equation}

Upon differentiating the RT formula with respect to $\ell$ and solving for $\ell$ in terms of $z_\ast$, the entanglement entropy can be rewritten as
\begin{equation}
S_{\text{strip}}(z_\ast) = \frac{\sqrt{2}}{3}\frac{ k^{1/2}N^{3/2}L}{a} - \frac{LR\pi^{3/2}k^{1/4}N^{3/4}2^{1/4}}{\sqrt{3G_{N}^{(4)}}\Gamma(1/4)^2z_\ast}.
\end{equation}
The RT formula then gives us the integral equation $S_{\text{strip}}(z_\ast) = \mathcal{A}(z_\ast)/4G^{(4)}_N$ for the function $f(z)$. The details of solving the integral equation are presented in Appendix A. The result is
\begin{equation}
\lim_{a/z \to 0} f(z)^2 = \frac{G_{N}^{(4)}k^{1/2}N^{3/2}2\sqrt{2}}{3R^2}.
\end{equation}
We are only interested in the regime $a/z \ll 1$ because it corresponds to the UV cutoff becoming small and the metric not probing too close to the boundary at $z = 0$.

After rescaling $z$, the metric is
\begin{equation}
ds^2 = \frac{\tilde{R}^2}{\tilde{z}^2}\Big(-h(z)dt^2 + dz^2 + dx^2 + dy^2\Big),
\end{equation}
where the AdS radius is
\begin{equation}
\tilde{R} = \left(\frac{G_{N}^{(4)}k^{1/2}N^{3/2}2\sqrt{2}}{3}\right)^{1/2}.
\end{equation}
As a sanity check, notice that from \cite{Ryu:2006ef} we find
\begin{equation}
G_{N}^{(4)} = \frac{48\pi^{3}\ell_{p}^{9}}{(2R_{AdS_{4}})^{7}}, \quad
2R_{AdS_{4}} = \ell_{p}(32\pi^{2}k^{1/3}N)^{\frac{1}{6}}.
\end{equation}
Solving these two gives us
\begin{equation}
R_{AdS_{4}}^{2} = \frac{G_{N}^{(4)}k^{1/2}N^{3/2}2\sqrt{2}}{3},
\end{equation}
which is exactly what we have found!

\subsection{Fixing the last metric component}

\indent The time-independent RT prescription is only enough to determine one of the unknown functions in the metric ansatz. With our assumptions about the holographic dictionary at the beginning of section 4 and some mild assumptions about the entanglement entropy in ABJM, it is safe to assume (see for example \cite{Faulkner:2017tkh}) that the bulk metric satisfies Einstein's equations.

Using $\Lambda = -3/\tilde{R}^2$, the three unique Einstein's equations are
\begin{align}
\frac{12h}{z^2} + 2 h'' - \frac{(h')^2}{h} - \frac{6h'}{z^2} &= \frac{12}{z^2} \\
\frac{\frac{z\Big(2h\big(h' - 2h'') + z(h')^2\Big)}{h^2} - 12}{4z^2} &= -\frac{3}{z^2} \\
\frac{\frac{zh'}{h} - 6}{2z^2} &= -\frac{3}{z^2}.
\end{align}
The final one implies $h = $ const. while the first two are consistent only if $h = 1$. Thus we have recovered the metric of $AdS_4$
\begin{equation}
ds^2 = \frac{\tilde{R}^2}{z^2}(-dt^2 + dz^2 + dx^2 + dy^2)
\end{equation}
with $\tilde{R}$ given above.

\section{Discussion}

\indent In this paper we have demonstrated again the relevance of entanglement entropy to the structure of spacetime. In particular, we have used general ideas from holography to reconstruct the metric of the bulk spacetime dual to ABJM in its ground state in flat space. This result enlarges the list of AdS/CFT pairs that have been investigated through this specific methodology.

It is worth noting an apparent tension between our philosophy and the actual source of the entanglement entropy in eqn. (\ref{eqn:Sstrip}). This result was derived in the original paper by Ryu and Takayanagi \cite{Ryu:2006bv}. That is to say, they predicted the entanglement entropies for a theory dual to $AdS_4 \times S^7$ by doing a calculation in the bulk (they did not know that they were computing the entanglement for ABJM with $k = 1$ because the RT paper predated the original ABJM paper \cite{Aharony:2008ug}). There appears to be a circularity in using a bulk calculation to reconstruct the bulk metric. However, there is a good reason to overlook that. It was shown by direct computation on the field theory side \cite{Drukker:2010nc, Klebanov:2011td} that the RT prediction for a spherical entangling surface can be trusted. Moreover, the authors of \cite{Saha:2018jjb} used a result from RT \cite{Ryu:2006bv} to reconstruct the bulk metric dual to $\mathcal{N} = 4$ SYM in flat space. 

It would be interesting to test the methodology of this paper with a time dependent entanglement entropy in the CFT. We suspect that the covariant generalization of the RT prescription \cite{Hubeny:2007xt} would be necessary. In the static examples considered here and in \cite{Saha:2018jjb} the time component of the metric is determined by demanding that Einsten's equations be satisfied, {\it i.e.} not using the entanglement entropy. It would further probe the connection between entanglement and spacetime by trying to fully reconstruct a dynamic bulk metric using a dynamic entanglement entropy.

\section*{Acknowledgements}
\noindent This work is supported by the DOE grant DE-SC0011687. We would like to thank Felipe Rosso, Clifford Johnson, and Kryzystof Pilch for helpful discussions.

\section*{Appendix A}
We now begin to solve the integral equation for the function $f$. We write
\begin{equation}
\mathcal{F}(z_\ast) = \frac{4G_{N}^{(4)}S_{EE}(z_\ast)}{2LR^2 z_\ast^2}, \quad p(z) = \frac{f(z)}{z^2}, \quad g(z) = z^4
\end{equation}
so that
\begin{equation}
\mathcal{F}(z_\ast) = \int_a^{z_\ast}dz\, \frac{p(z)}{\sqrt{g(z_\ast) - g(z)}}.
\end{equation}
This integral equation has the solution
\begin{equation}
p(z) = \frac{1}{\pi} \frac{d}{dz} \int_a^z dz_\ast \, \frac{\mathcal{F}(z_\ast) g'(z_\ast)}{\sqrt{g(z) - g(z_\ast)}},
\end{equation}
where $g'(z) = \frac{dg}{dz} > 0$. So, we find
\begin{equation}
f(z) = \frac{8G_{N}^{(4)}z^2}{\pi LR^2} \frac{d}{dz} \int_a^z dz_\ast \, \frac{z_\ast S_{EE}(z_\ast)}{\sqrt{z^4 - z_\ast^4}}.
\end{equation}
Lets us define:
\begin{align}
C_1 &= \frac{\sqrt{2}}{3}\frac{ N^{3/2}L}{a}, \\
C_2 &=  \frac{LR\pi^{3/2}N^{3/4}2^{1/4}}{\sqrt{3G_{N}^{(4)}}\Gamma(1/4)^2z_\ast}.
\end{align}
Thus
\begin{equation}
f(z) = \frac{8G_{N}^{(4)}z^2}{\pi LR^2} \frac{d}{dz} \int_a^z dz_\ast \, \frac{C_1 z_\ast - C_2}{\sqrt{z^4 - z_\ast^4}}  \equiv \frac{8G_{N}^{(4)}z^2}{\pi L R^2} \frac{dM}{dz}.
\end{equation}
One finds
\begin{equation}
M(z) = C_1\left[ \frac{\pi}{4} - \frac{1}{2} \tan^{-1}\left(\frac{a^2}{\sqrt{z^4 - a^4}}\right)\right] - C_2\left[\frac{\sqrt{\pi}\,\Gamma\left(\frac{5}{4}\right)}{z\Gamma\left(\frac{3}{4}\right)} - \frac{a\, _2F_1\left(\half, \frac{1}{4}, \frac{5}{4}, \frac{a^4}{z^4}\right)}{z^2}\right],
\end{equation}
where $_2F_1$ is a hypergeometric function. Therefore
\begin{equation}
z^2 \frac{dM}{dz} = \frac{C_1 a^2}{z\sqrt{1 - \frac{a^4}{z^4}}} + \frac{C_2 \sqrt{\pi}\Gamma\left(\frac{5}{4}\right)}{\Gamma\left(\frac{3}{4}\right)} - \frac{C_2 a}{z}\left(\frac{1}{\sqrt{1 - \frac{a^4}{z^4}}} + \,_2F_1\left(\frac{1}{4}, \half, \frac{5}{4}, \frac{a^4}{z^4}\right)\right).
\end{equation}
In the limit $a/z \to 0$ we have
\begin{equation}
\lim_{a/z \to 0} \frac{8G_{N}^{(4)}z^2}{\pi L R^2} \frac{dM}{dz} = \frac{C_2 \sqrt{\pi}\Gamma\left(\frac{5}{4}\right)}{\Gamma\left(\frac{3}{4}\right)} = \text{const.}
\end{equation}
Using the expression of $C_2$ and the fact that
\begin{equation}
\Gamma\left(\frac{5}{4}\right) = \frac{\Gamma\left(\frac{1}{4}\right)}{4},
\end{equation}
we end up with
\begin{equation}
\lim_{a/z \to 0} f(z)^2 = \frac{G_{N}^{(4)}N^{3/2}2\sqrt{2}}{3R^2}.
\end{equation}

\bibliography{abjmpaper}
\bibliographystyle{hephys}

\end{document}